\documentclass[12pt]{article}
\usepackage{latexsym,epsfig,amssymb,amsmath,cite,verbatim}

\newcommand{\cM}{{\cal M}}
\newcommand{\cN}{{\cal N}}

\newcommand{\mZ}{\mathbb{Z}}
\newcommand{\mI}{\mathbb{I}}
\newcommand{\mR}{\mathbb{R}}
\newcommand{\nH}{n_{\rm H}}
\newcommand{\nV}{n_{\rm V}}

\def\bfone{\relax{\rm 1\kern-.35em 1}}

\newcommand{\be}{\begin{equation}}
\newcommand{\ee}{\end{equation}}
\newcommand{\ben}{\begin{displaymath}}
\newcommand{\een}{\end{displaymath}}
\newcommand{\bea}{\begin{eqnarray}}
\newcommand{\eea}{\end{eqnarray}}

\newcommand{\bean}{\begin{eqnarray*}}
\newcommand{\eean}{\end{eqnarray*}}

\makeatletter \@addtoreset{equation}{section} \makeatother

\begin{document}

\begin{titlepage}

\begin{flushright}
\small ~~
\end{flushright}

\bigskip

\begin{center}

\vskip 2cm

{\LARGE \bf De Sitter in Extended Supergravity} \\[6mm]

{\bf Diederik Roest and Jan Rosseel}\\

\vskip 25pt

{\em Centre for Theoretical Physics,\\
University of Groningen, \\
Nijenborgh 4, 9747 AG Groningen, The Netherlands\\
{\small {\tt \{d.roest, j.rosseel\}@rug.nl}}} \\

\vskip 0.8cm

\end{center}

\vskip 1cm

\begin{center} {\bf ABSTRACT}\\[3ex]

\begin{minipage}{13cm}
\small

We show that known de Sitter solutions in extended gauged supergravity theories are interrelated via a web of supersymmetry-breaking truncations. In particular, all $\cN = 8$ models reduce to a subset of the $\cN = 4$ possibilities. Furthermore, a different subset of the $\cN = 4$ models can be truncated to stable de Sitter vacua in $\cN = 2$ theories. In addition to relations between the known cases, we also find new (un)stable models.

\end{minipage}

\end{center}


\vfill

\end{titlepage}



\section{Introduction}


In $\cN = 1$ supergravity theories, one has a fair amount of freedom in introducing a scalar potential. Starting from an ungauged or massless theory, one can introduce a holomorphic superpotential. This situation radically changes when considering extended, i.e.~$\cN \geq 2$, supergravities. The scalar potential of extended theories is fully determined by gaugings: one cannot introduce a scalar potential without turning on a gauging, and similarly one cannot turn on a gauging without inducing a scalar potential. One might think that in $\cN =2$ theories without hypermultiplets there is the additional possibility of introducing Fayet-Iliopoulos terms, but in fact these can also be understood as a gauging of the $SU(2)$ R-symmetry group \cite{toine}. Therefore a scalar potential can only be obtained by introducing a number of constants that specify the gauging, instead of using an arbitrary $\cN =1$ holomorphic superpotential. For this reason the scalar potentials of extended gauged supergravity have a certain rigidity, and one may wonder to what extent these allow for cosmologically interesting solutions, e.g.~stable de Sitter solutions or slow-roll inflationary models.

Starting from the theory with maximal supersymmetry, for particular gaugings the $\cN = 8$ scalar potential does indeed allow for stationary points with a positive scalar potential \cite{HW84}. However, in all known cases these are saddlepoints rather than minima, and hence the corresponding de Sitter solution is unstable. In fact, the second slow roll parameter $\eta$, which is defined as the lowest eigenvalue of the scalar mass matrix divided by the value of the scalar potential in the extremum,
 \begin{align}
  m_{ij} = V^{-1} D_i \partial_j V  \,,
 \end{align}
where $D_i$ is the covariant derivative on the scalar manifold, always takes the value $\eta = -2$ \cite{Kallosh}. Therefore these unstable de Sitter solutions are also unsuitable for slow-roll inflation, as this requires $|\eta| \ll 1$.

The situation ameliorates somewhat when going down to $\cN = 4$ supersymmetries. A systematic classification of semi-simple gauge groups giving rise to de Sitter vacua was performed for $\cN = 4$ theories with six vectormultiplets in \cite{Westra1, Westra2}.  It turns out to be possible to raise the value of $\eta$ above $-2$; however, it always remains negative. Again stable de Sitter and/or slow-roll inflation is impossible in all known cases.

Things become more interesting when considering $\cN =2$ theories. The implications of this smaller amount of supersymmetries differ from the previous situations in a number of respects: one can introduce both vector- and hypermultiplets and the scalar manifolds are no longer uniquely determined by the matter content. Indeed it turns out to be possible to evade the no-go theorems of e.g.~\cite{Cremmer, GomezReino} and to construct models with (meta-)stable de Sitter vacua, i.e.~with $\eta$ vanishing or positive \cite{FTP}. In these models, the gauge groups consist of two factors: a non-compact electric factor and a compact magnetic factor. Crucially, the compact factor should have a non-trivial action on the hypersector, or Fayet-Iliopoulos terms in the absence of hypermultiplets.

The different gaugings for the various amounts of supersymmetry have been constructed independently and may seem unrelated. However, it is the purpose of this paper to show that they are in fact interrelated via a web of supersymmetry truncations. Indeed, all known $\cN = 8$ gaugings with de Sitter solutions can be related to a subset of the $\cN = 4$ models. Moreover, one can construct all known $\cN =2$ models with stable de Sitter solutions from a different subset of the $\cN = 4$ models. This offers a unifying picture of all known gaugings of extended supergravity with (un)stable de Sitter solutions. In the process of showing this, we will also construct new unstable $\cN = 4$  and stable $\cN =2$ models. 

This paper is organised as follows. In section 2 we will review and slightly generalise the $\cN =4$ gaugings. Their relation to $\cN = 8$ gaugings is discussed in section 3. Subsequently, we truncate to stable $\cN = 2$ gaugings in section 4. Finally, section 5 contains our conclusions and outlook.


\section{The $\cN = 4$ models revisited}


Our starting point will be the possible gaugings of $\cN =4$ gauged supergravity. A thorough discussion of this theory can be found in e.g.~\cite{SW}, and we will only present the necessary details and formulae here. We will restrict to the case of six vectormultiplets, as this is most relevant for the discussion of de Sitter vacua in the literature. In this case the global symmetry is $SL(2) \times SO(6,6)$. The scalars parametrise the corresponding cosets of dimensions 2 and 36, while the 12 vectors transform in the fundamental representation of $SO(6,6)$. Furthermore, the electric and magnetic parts of all vectors form a doublet under $SL(2)$. 

As for gaugings, it was shown in \cite{SW} that these are parametrised by components $f_{\alpha MNP}$ and $\xi_{\alpha M}$, where $\alpha = (\uparrow,\downarrow)$ are $SL(2)$ indices and $M = (1, \ldots, 6, \bar 1, \ldots \bar 6)$ are $SO(6,6)$ light-cone indices. Components with indices $\alpha = \uparrow$ correspond to gaugings employing the electric part of the vectors, while components with $\alpha = \downarrow$ pertain to magnetic gaugings. Consistency of the gaugings imposes a number of quadratic constraints on these components, corresponding to Jacobi identities and orthogonality of charges. Together with the explicit form of the scalar potential these can be found in \cite{SW}.

We will first restrict to a subset of all possible gaugings, namely those that correspond to a direct product of two six-dimensional gauge factors, $G = G_1 \times G_2$, embedded in an $SO(3,3)^2$ subgroup of the global symmetry group. This implies that $\xi_{\alpha M} = 0$, as these components induce gaugings of (subgroups of) $SL(2)$. Furthermore, only 40 of the 220 doublet components of $f_{\alpha MNP}$ survive this truncation. By performing $SL(2)$ transformations we can subsequently arrange $G_1$ to be an electric gauging, while $G_2$ is magnetic\footnote{In particular, an $SO(2)$ rotation can be used to bring e.g.~$G_1$ to a purely electric gauging. Subsequently we can use the shift symmetry of the axion to make $G_2$ purely magnetic, provided it had a magnetic component to start with \cite{DR}.}. It will be useful to organise the remaining 40 components in four symmetric matrices with $SL(4)$ indices $i = 1, \ldots, 4$. To this end we relate the anti-symmetric representation of $SL(4)$ to the fundamental of $SO(6,6)$ via
 \begin{align}
  (1,2,3,\bar 1, \bar 2, \bar 3) \simeq (12,13,14,43,24,32) \,.
 \end{align}
This allows us to express the structure constants in terms of charge matrices $Q_1$, ${\tilde Q}_1$:
 \begin{align}
   f_{\uparrow ij,kl}{}^{mn} = 8 \delta_{[i}^{[m} Q_{1j][k} \delta_{l]}^{n]} - \epsilon_{iji'j'} \epsilon_{klk'l'} \epsilon^{mnm'n'} \delta^{i'}_{m'} {\tilde Q}_1^{j'k'} \delta^{l'}_{n'} \,. \label{strconst}
 \end{align}
Restricting to diagonal matrices this reduces to
  \begin{align}
   Q_1 = \text{diag} ( f_{\uparrow 123}, f_{\uparrow 1 \bar 2 \bar 3}, f_{\uparrow \bar 1 2 \bar 3}, f_{\uparrow \bar 1 \bar 2 3} ) \,, \quad
   {\tilde Q}_1 = \text{diag} ( f_{\uparrow \bar 1 \bar 2 \bar 3}, f_{\uparrow \bar 1  2  3}, f_{\uparrow 1  \bar 2  3}, f_{\uparrow 1 2 \bar 3} ) \,.
  \end{align}
Similar expressions relate the structure constants of the second $SO(3,3)$ factor to two matrices $Q_2$ and ${\tilde Q}_2$.

For this subset of gaugings, one can check that the general form of the scalar potential $V$ as given in \cite{SW}, restricted to the dilatons, can be written in terms of a superpotential $W$:
 \begin{align}
  V = \tfrac12 (\partial_{\vec \phi} W)^2 - \tfrac38 W^2 \,, \quad
  W = e^{\phi /2} {\rm Tr}[Q_1 \cM_1 - \tilde{Q}_1 \cM_1^{-1}] + e^{- \phi /2} {\rm Tr}[Q_2 \cM_2 - \tilde{Q}_2 \cM_1^{-2}] \,,
 \end{align}
where $\vec \phi = (\phi, \phi_1, \ldots, \phi_6)$. The first of these corresponds to the $SL(2)$ factor, while the $SO(6,6)$ dilatons are parametrised by two $SL(4)$ factors of the form
 \begin{align}
 \cM_1^{ij} = \text{diag} (e^{\vec{\alpha}_1 \cdot \vec{\phi}},\ldots,e^{\vec{\alpha}_4 \cdot \vec{\phi}}) \,,
 \end{align}
where the $4$ vectors $\vec{\alpha}_i=\{\alpha_{iI}\}$ are weights of $SL(4,\mathbb{R})$:
 \begin{align}
  \alpha_{iI} = \frac{1}{\sqrt{2}} \left( \begin{array}{ccc} 1 & 1 & 1 \\ 1 & -1 & -1 \\ -1 & 1 & -1 \\ -1 & -1 & 1 \end{array} \right) \,.
 \end{align} 
with $I = 1, \ldots, 3$. The definition for $\cM_2$ in terms of $\phi_{4,5,6}$ is analogous.

The four matrices describe gaugings of $CSO(p,q,4-p-q)$ in either of the two $SO(3,3) \simeq SL(4,\mathbb{R})$ factors\footnote{In general the embedding of $CSO(p,q,n-p-q)$ in $SL(n)$ is specified by a single matrix $Q$ (see e.g.~the next section for $n=8$). Due to the isomorphism $SL(4) \simeq SO(3,3)$ for $n=4$, there is an additional invariant tensor (the Levi-Civita symbol) to construct more general structure constants in \eqref{strconst} and hence one needs two matrices.}. To see this, let us look at the first $SO(3,3)$ factor, spanned by $(1,2,3,\bar 1, \bar 2, \bar 3)$, in detail. The formulae are completely analogous for the second factor. Restricting to (semi-)simple gaugings\footnote{Non-semi-simple gaugings correspond to having the determinant of both $Q_1$ and $\tilde Q_1$ vanishing. In this way one can make contact with the $CSO$-gaugings considered in \cite{CSO}.}, the Jacobi identities on the structure constants \eqref{strconst} imply that the matrices $Q_1$ and $\tilde{Q}_1$ are proportional. This leads to three physically distinct possibilities:
 \begin{align}
  Q_1 = g_1 \eta \,, \qquad {\tilde Q}_1 = {\tilde g}_1 \eta \,, \qquad \eta = 
 \begin{cases} 
  \text{diag}(+1,+1,+1,+1) \quad - \; SO(4) \,, \notag \\
  \text{diag}(+1,+1,+1,-1) \quad - \; SO(3,1) \,, \notag \\
  \text{diag}(+1,+1,-1,-1) \quad - \; SO(2,2) \,, \notag \\  
 \end{cases}
 \end{align}
where $\eta_{ij}$ is the invariant metric of the gauge group. 

To make this more explicit, let us write out the structure constants in Cartesian coordinates with $SO(6,6)$ metric $\eta_{MN} = \text{diag}(-1,\ldots,-1,1,\ldots,1)$. For $SO(4)$ gaugings we find
 \begin{align}
  f_{\uparrow 123} = \sqrt{2} (g_1 - \tilde g_1) \,, \quad f_{\uparrow 789} = \sqrt{2} (g_1 + \tilde g_1) \,.
 \end{align}
For $SO(3,1)$ gaugings we find
  \begin{align}
  & f_{\uparrow 123} = - f_{\uparrow 783} = f_{\uparrow 729} = f_{\uparrow 189} = \tfrac12 \sqrt{2} (g_1 - \tilde g_1) \,, \notag \\
  & f_{\uparrow 789} = - f_{\uparrow 129} = f_{\uparrow 183} = f_{\uparrow 723} = \tfrac12 \sqrt{2} (g_1 + \tilde g_1) \,. \label{SO31}
 \end{align}
Finally, for $SO(2,2)$ gaugings we find
 \begin{align}
  f_{\uparrow 723} = \tfrac12 \sqrt{2} (g_1 + \tilde g_1) \,, \quad f_{\uparrow 189} = \tfrac12 \sqrt{2} (g_1 - \tilde g_1) \,.
 \end{align}
These gaugings have been previously considered in \cite{Westra2}. Concerning the two non-simple cases, one finds that both simple factors are non-vanishing for generic values of $g_1$ and $\tilde g_1$. The singular cases arise when either of the two simple factors vanish, i.e.~$g_1 = \pm \tilde g_1$. The $SO(3,1)$ gauging is somewhat more intricate. For special values of $g_1$ and $\tilde g_1$, it corresponds to either of the embeddings of \cite{Westra2}: for $g_1 = \pm \tilde g_1$ it is the $SO(3,1)_\pm$ embedding. For these embeddings, the compact generators are given by either $T_{7,8,9}$ or $T_{1,2,3}$, respectively. However, we find a one-parameter family of different embeddings, labelled by e.g.~$\tilde g_1 /  g_1$. Another interesting special case is $\tilde g_1 = 0$, which we will refer to as the null embedding, denoted by $SO(3,1)_0$. For this embedding the compact generators are null linear combinations of the generators $T_{1,2,3}$ and $T_{7,8,9}$. 

To summarise the discussion so far, restricting to the (semi-)simple cases, there are three inequivalent gaugings in the first $SO(3,3)$ factor, all specified by two parameters. In total there are six inequivalent gaugings, specified by four parameters:
 \begin{align}
  \left( \text{electric} \begin{cases}
   SO(4)  \\ SO(3,1)  \\ SO(2,2)  
  \end{cases} \text{with }g_1, \tilde g_1 \right) \; \;
  \times_{\text{symm}}  
  \left( \text{magnetic}
  \begin{cases}
   SO(4)  \\ SO(3,1)  \\ SO(2,2)
  \end{cases} \text{with }g_2, \tilde g_2 \right) \,.
   \label{6Dgg}
 \end{align}
In \cite{Westra2} it has been analysed which of these give rise to de Sitter vacua, with the restriction to the $SO(3,1)_\pm$ embeddings. All de Sitter vacua turned out to have an instability, either in the $SL(2)$ or in the $SO(6,6)$ part. We will not analyse the remaining possibilities, with other $SO(3,1)$ embeddings, in detail at this point. However, we have looked at a few possibilities and found that the resulting de Sitter vacua are unstable as well. One interesting point is that the scalar mass matrix no longer splits up in an $SL(2)$ and an $SO(6,6)$ part in general, as was the case for \cite{Westra2}.

A number of generalisations is possible at this point \cite{Westra2}. First of all, in the case of non-simple gauge groups in \eqref{6Dgg}, one can choose a different $SL(2)$ angle for the different simple factors, allowing in total up to four angles. Secondly, one can think about other real forms of the $(A_1)^4$ gauge group. It turns out that the only additional possibility, not of the form \eqref{6Dgg}, is given by $SO(3) \times SO(2,1)^3$. Finally, the remaining option is to replace three $A_1$ factors by an eight-dimensional $A_2$ factor. This leads to e.g.~the possibility $SU(2,1) \times SO(2,1)$.

In the following sections we will show how all known $\cN=8$ and $\cN=2$ gauged supergravity models with de Sitter vacua are related to this simple set of $\cN=4$ models. 


\section{View from the top: from $\cN = 8$ to $\cN = 4$}


An important class of gaugings in $\cN = 8$ supergravity are the $SO(p,8-p)$ gaugings, or their contractions $CSO(p,q,8-p-q)$. These gauge a subgroup in the manifest $SL(8,\mR)$ subgroup\footnote{There are other formulations or duality frames of $\cN = 8$ supergravity, where a different subgroup of $E_{7(7)}$ is manifest, i.e.~realised on the electric vectors. An example is $SL(3) \times SL(6)$. It would be of interest to see if this formulation allows for gaugings that might reduce to the more exceptional gaugings of $\cN = 4$ supergravity with a 9+3 split.} of the hidden symmetry group $E_{7(7)}$. The $CSO(p,q,8-p-q)$ gaugings are determined by a symmetric matrix $Q_{ij}$. By choosing an appropriate basis this matrix can always be diagonalised with diagonal entries equal to $0$ or $\pm 1$. The number of positive entries corresponds to $p$ and the negative ones to $q$. The remaining $8-p-q$ entries vanish. 

The scalar potential $V$ is given in terms of a superpotential $W$:
 \begin{align}
  V = \tfrac12 (\vec{\partial}_\phi W)^2 - \tfrac38 W^2 \,, \quad W= \text{Tr}[Q \cM] \,,
 \end{align}
where the scalars in the $SL(8, \mathbb{R}) / SO(8)$ part of the $E_{7(7)} / SU(8)$ coset are parametrised by a scalar matrix $\cM^{ij}$. The dilatons are the diagonal part of this scalar matrix,
 \begin{align}
  \cM = \left( \begin{array}{cc} e^{\phi /2} \cM_1 & \\ & e^{-\phi /2} \cM_2 \end{array} \right) \,,
 \end{align}
where $\cM_{1,2}$ have been defined in the previous section.

These gaugings give rise to Anti-de Sitter vacua for the $SO(8)$ gauging, and de Sitter vacua for $SO(5,3)$ and $SO(4,4)$ \cite{HW84}. For the intermediate cases $SO(7,1)$ and $SO(6,2)$ there is no maximally symmetric vacuum. It can be checked that the dilatons of the two de Sitter vacua always contain one tachyonic direction. In particular, the second derivatives of the scalar potential have the following eigenvalues in the vacuum:
 \begin{align}
  \text{eigenvalues~} (V^{-1} \partial_{\vec \phi} \partial_{\vec \phi} V) & = (-1,1,1,1,1,1,1) \,, \quad \text{for }SO(4,4) \,, \notag \\
   & = (-1, \tfrac23, \tfrac23, \tfrac23, \tfrac23, 2, 2) \quad \text{for }SO(5,3) \,.
 \end{align}
  
To go down to $\cN = 4$ supersymmetry one mods out the $\cN = 8$ theory by a particular $\mathbb{Z}_2$ element of the R-symmetry group $SU(8)$:
 \begin{align}
  \left( \begin{array}{cc} \mI_4 & \\ & - \mI_4 \end{array} \right) \,.
 \end{align}
This leaves half of the gravitini and the supersymmetries invariant and projects out the other half. For the vectors, it leaves 12 out of 28 invariant. The scalar coset $SL(8, \mathbb{R}) / SO(8)$ reduces to $\mR^+ \times SL(4, \mathbb{R}) / SO(4) \times SL(4, \mathbb{R}) / SO(4)$. As $Q$ is diagonal, this is always invariant under this truncation. In other words, all truncations of these $\cN =8$ gaugings to $\cN = 4$ are consistent.

\begin{table}[bt]
\begin{center}
\begin{tabular}{||lcl||}
\hline 
$\cN = 8$ gauging & $\rightarrow$ & $\; \cN =4$ gauging \\ \hline \hline
$SO(4,4)$ & $\rightarrow$ & $\begin{array}{l} SO(4) \times SO(4) \\ SO(3,1) \times SO(3,1) \\ SO(2,2) \times SO(2,2) \end{array}$ \\ \hline
$SO(5,3)$ & $\rightarrow$ & $\begin{array}{l} SO(4) \times SO(3,1) \\ SO(3,1) \times SO(2,2) \end{array}$ \\ \hline
\end{tabular}
\end{center}
\caption{Truncations of $\cN =8$ gaugings with unstable de Sitter solutions to $\cN = 4$ gaugings. The resulting theories always have two factors with orthogonal angles and null embeddings, i.e.~$\tilde{g}_1 = \tilde{g}_2 = 0$. All de Sitter solutions remain unstable in $\cN = 4$.} \label{tab:N=8trunc}
\end{table}

As can easily be seen from a comparison of the superpotentials, the truncation of the $\cN = 8$ gaugings lead to the subset of $\cN = 4$ models that have\footnote{The truncation of the $SO(8)$ gauged theory to $SO(4)^2$ in $\cN = 4$ was already pointed out in \cite{HW85}.}
 \begin{align}
  Q = \left( \begin{array}{cc} Q_1 & \\ & Q_2 \end{array} \right) \,,
 \end{align}
and vanishing $\tilde Q$-matrices. The resulting $\cN = 4$ $SL(2)$ angles are completely fixed: the first factor is electric while the second is magnetic. We therefore end up with exactly the set of gaugings given in \eqref{6Dgg} with $\tilde{g}_1 = \tilde{g}_2 = 0$. Of these, the set of gaugings that come from $SO(4,4)$ or $SO(5,3)$, which are five of the possible combinations in \eqref{6Dgg}, have a de Sitter solution. Only the $SO(4) \times SO(2,2)$ gauging follows from $SO(6,2)$ and does not have a maximally symmetric solution. 


\section{Descending into stability: from $\cN =4$ to $\cN = 2$}


In this section, we will discuss truncations of the $\mathcal{N}=4$ gauged supergravities with de Sitter vacua, as identified in \cite{Westra1, Westra2}, that break supersymmetry down to $\cN = 2$. Our goal is to obtain $\mathcal{N}=2$ theories where the de Sitter vacuum is stable. Since models with tachyonic $SL(2)$ scalars do not lead to stable $\mathcal{N}=2$ de Sitter vacua, for reasons that will become clear later on, we will discard these from now on. This leaves us with five possible gauge groups, that can be found in e.g.~table \ref{tab:2244trunc}.

In general, truncations to $\cN = 2$ are achieved by modding out with respect to a $\mZ_2$ element of $SO(6) \times SO(6) \subset SO(6,6)$ of the form\footnote{The form of this $\mathbb{Z}_2$-truncation is defined up to permutations of the diagonal elements.}
 \begin{align} \label{z2}
   \left( \begin{array}{cc} + \mI_2 & \\ & - \mI_4 \end{array} \right) \times 
   \left( \begin{array}{cc} + \mI_{n_{\rm V} -1}   & \\ & - \mI_{n_{\rm H}} \end{array} \right) \,,
 \end{align}
with $\nV + \nH = 7$. The spectrum of fields that are even under this $\mathbb{Z}_2$ truncation, is that of an $\mathcal{N}=2$ supergravity, with $n_V$ vectormultiplets and $n_H$ hypermultiplets. The scalar fields then span the following symmetric scalar manifold\footnote{Amusingly, it has been argued in \cite{FTP} that these particular $\cN = 2$ scalar manifolds are the only ones that can allow for stable de Sitter solutions.}:
\begin{align}
  \cM = \underbrace{\frac{SL(2)}{SO(2)} \times \frac{SO(2,\nV -1)}{SO(2) \times SO(\nV -1)}}_{\rm SK} \times \underbrace{\frac{SO(4, \nH)}{SO(4) \times SO(\nH)}}_{\rm QK} \,. \label{Z2}
 \end{align}
The first two factors span a special K\"{a}hler manifold, while the third factor corresponds to a quaternionic-K\"{a}hler space. The $SO(6,6)$ part of the $\cN = 4$ scalar manifold is truncated into two $SO(p,q)$ parts, while the $SL(2)$ part remains intact. This is the reason for discarding the $\cN = 4$ models with unstable $SL(2)$ scalars: upon truncation, the instability is inherited by the $\cN = 2$ theory.

In order for the truncation (\ref{z2}) to be consistent in the presence of a gauging, the structure constants of the gauge group have to be even under the $\mZ_2$ truncation. The $\mathcal{N}=4$ gaugings in general then lead to $\mathcal{N}=2$ gaugings and possible Fayet-Iliopoulos terms. Which $\mathcal{N}=2$ gauging one is left with and whether or not the tachyonic scalars are truncated, has to be checked in all different cases. 

For reasons of clarity, we will first discuss an example of such a truncation in more detail, before tackling the general case. Consider the gauge group $SO(3,1) \times SO(3,1)$. We will embed the adjoint of the first $SO(3,1)$ factor along the indices $1, 2, 3, 7, 8, 9$, and the adjoint of the second factor along the directions $4, 5, 6, 10, 11, 12$. Furthermore, the rotation subgroups of the two $SO(3,1)$ subgroups lie along the indices $7, \ldots, 12$, the boosts along the indices $1\ldots 6$. The structure constants corresponding to the first gauge factor can be read off from \eqref{SO31} with $g_1 = \tilde{g}_1$, and similar for the second gauge factor. Consider then the $\mZ_2$ truncation
 \begin{align} 
   \left( \begin{array}{cc} + \mI_2 & \\ & - \mI_4 \end{array} \right) \times 
   \left( \begin{array}{cc} - \mI_2 & \\ & + \mI_4  \end{array} \right) \,.  \label{Z42}
 \end{align}
The structure constants \eqref{SO31} are even under this $\mathbb{Z}_2$ element and the resulting $\mathcal{N}=2$ supergravity has $\nV = 5$ vectormultiplets and $\nH = 2$ hypermultiplets. The resulting $\mathcal{N}=2$ gauge group is given by $SO(2,1) \times SO(3)$. The first factor is spanned by the gauge vectors in the $(1,2,9)$-directions, while the second factor is spanned by the $(10,11,12)$ gauge vectors. Crucially, both gauge factors act in both the special K\"{a}hler $SO(2,4)$ and in the quaternionic-K\"{a}hler $SO(4,2)$ part, as can be seen from the adjoint representation of these generators. This truncation corresponds to the third model of \cite{FTP} with $r_0 = 1$. This identification is confirmed by looking at the value of the scalar potential and its second derivatives. From \cite{Westra2} one finds that the value of the scalar potential in the de Sitter extremum is
 \begin{align}
  V_0 = 3 | g_1 g_2 \sin(\alpha_1 - \alpha_2)| \,,
 \end{align}
where $\alpha_1 - \alpha_2$ is the $SL(2)$ angle between the two $SO(3,1)$ factors. This exactly coincides with equation (3.47) of \cite{FTP} for $r_0 = 1$. Similarly, the $\cN = 4$ eigenvalues of the scalar mass matrix, divided by $V_0$, are
 \begin{align}
   - \tfrac43 \, (1\times), \, 0 \, (15\times), \, \tfrac23 \, (10\times), \, \tfrac43 \, (9\times), \, \tfrac83 \, (1\times), \, 2 \, (2\times), \, 
 \end{align}
where the latter correspond to the $SL(2)$ scalars. We have explicitly checked that these are truncated to the $\cN = 2$ subset
 \begin{align} \label{truncexmasses}
   0 \, (8\times), \, \tfrac23 \, (2\times), \, \tfrac43 \, (6\times), \, 2 \, (2\times), \, 
 \end{align}
which again coincides with table 2 of \cite{FTP}. 

\begin{table}[bt]
\begin{center}
\begin{tabular}{||lclc||}
\hline
$\mathcal{N}=4$ gauging & $\rightarrow$ & $\; \mathcal{N}=2$ gauging with $n_V = 3$, $n_H = 4$ & Stable? \\ \hline \hline
$SO(3,1)_+ \times SO(3,1)_+$ & $\rightarrow$ & $\; (SO(2)_+ \bowtie SO(1,1)_-)^2$ & $-$ \\ \hline
$SO(3,1)_+ \times SO(2,2)_\pm$ & $\rightarrow$ & $\begin{array}{l} SO(2,1)_+^H \times SO(2)_+ \\
SO(2)_+ \bowtie SO(1,1)_- \times SO(1,1)_\pm^2 \\ SO(2)_+ \bowtie SO(1,1)_- \times SO(2)_\pm^2
\end{array}$ &$ \begin{array}{c} - \\ - \\ - \end{array}$ \\ \hline $SO(2,2)_\pm \times SO(2,2)_\pm$ & $\rightarrow$ & $ \begin{array}{l}
SO(2,1)_+ \times SO(2)_+ \\ SO(1,1)_\pm^2 \times SO(1,1)_\pm^2  \\ SO(2)_\pm^2 \times SO(1,1)_\pm^2 \\ SO(2)_\pm^2 \times SO(2)_\pm^2
\end{array}$ & $\begin{array}{c} \surd \\ - \\ - \\ - \end{array}$  \\ \hline
$SO(3)_+ \times SO(2,1)_+^3$ & $\rightarrow$ & $\begin{array}{l} SO(2,1)_+ \times SO(2)_+ \\ SO(2)_+^2 \times SO(1,1)_-^2 \end{array}$  & $\begin{array}{c} - \\ - \end{array}$ \\ \hline
$SU(2,1)_+ \times SO(2,1)_+$ & $\rightarrow$ & $\; SO(2,1)_+^H \times SO(2)_+$ & $-$ \\ \hline
\end{tabular}
\end{center}
\caption{Truncations of $\cN =4$ gaugings with de Sitter solutions having $SO(6,6)$ instabilities to $\cN = 2$ gaugings with $n_V = 3$, $n_H = 4$.} \label{tab:2244trunc}
\end{table}

The previous example clarifies and corroborates the truncation procedure. One can fix the $\mathbb{Z}_2$ truncation, according to the number of vector- and hypermultiplets one wants to end up with. One then writes a form of the structure constants of the $\mathcal{N}=4$ gauge group that is $\mathbb{Z}_2$ invariant. In the following we will discuss truncations to $\mathcal{N}=2$ theories with either $n_V = 3$, $n_H = 4$, or $n_V = 5$, $n_H = 2$. Exhaustive lists of such gaugings that stem from truncation of an $\mathcal{N}=4$ supergravity with an $SO(6,6)$ unstable de Sitter vacuum are given in tables \ref{tab:2244trunc} and \ref{tab:2442trunc}. We use the following notation:
 \begin{itemize}
  \item
   The subscripts $+$ or $-$ on abelian factors indicate whether the generators are in the positive or negative part of the special orthogonal metrics, i.e.~are spanned by $T_{7,\ldots,12}$ or $T_{1,\ldots,6}$, respectively. For non-abelian factors the same holds for the compact generators, while the non-compact ones are in the part with the opposite sign. Subscripts $\pm$ are used for squares consisting of both embeddings, e.g.~$SO(2,2)_\pm = SO(2,1)_+ \times SO(2,1)_-$.
 \item
  Abelian factors always act non-trivially on the scalars in the hypersector (or else they would not have any effect on the scalar potential). The representations are always the fundamentals, i.e.~as matrices. The only exception is denoted by $SO(2) \bowtie SO(1,1)$ and comes from the $\cN=4$ gauge factor $SO(3,1)$. It acts in the following way on the hypermanifold: two temporal and two spatial directions form $SO(2)$ doublets, while the $SO(1,1)$ acts as a boost on the two doublets.
  \item
  For non-abelian factors, the superscript $H$ indicates that it acts non-trivially on the hyperscalars. For all gauge factors coming from $\cN = 4$ special orthogonal gaugings, the $\cN =2$ representation is the fundamental. The only exception comes from the $\cN = 4$ gauge factor $SU(2,1)$, in which case $SO(2,1)$ acts in the five-dimensional symmetric traceless representation.
  \item
  In some cases one has to remove $\cN = 4$ gauge factors in order to be able to truncate to $\cN =2$, leading to less gauge factors on the right side of the table. Of course one should make sure that such singular limits of the gauge group do not affect the de Sitter solution. This is the case for all possibilities listed in the tables.
  \end{itemize}

\begin{table}[bt]
\begin{center}
\begin{tabular}{||lclc||}
\hline
$\mathcal{N}=4$ gauging & $\rightarrow$ & $\; \mathcal{N}=2$ gauging with $n_V = 5$, $n_H = 2$ & Stable? \\ \hline \hline
$SO(3,1)_+ \times SO(3,1)_+$ & $\rightarrow$ & $\; SO(2,1)_+^H \times SO(3)_+^H$ & $\surd$ \\\hline
$SO(3,1)_+ \times SO(2,2)_\pm$ & $\rightarrow$ & $\begin{array}{ll}
SO(2,1)_+^H \times SO(2)_+ \\ SO(3)_+^H  \times SO(2,1)_+ \\ SO(3)_+^H \times SO(1,1)_- \\ SO(2,1)_- \times SO(2)_+ \bowtie SO(1,1)_- \times SO(2)_+ 
\end{array}$ & $\begin{array}{c} - \\ \surd \\ - \\ - \end{array}$ \\\hline $SO(2,2)_\pm \times SO(2,2)_\pm$ & $\rightarrow$ & $\begin{array}{ll}
SO(2,1)_- \times SO(2)_+^2 \times SO(2)_- \\ SO(2,1)_- \times SO(1,1)_\pm^2 \times SO(2)_+ \\ SO(2,1)_+ \times SO(1,1)_+^2 \times SO(2)_+ \\ SO(1,1)_-^2 \end{array}$ & $\begin{array}{c} - \\ - \\ \surd \\ - \end{array}$ \\\hline $SO(3)_+ \times SO(2,1)_+^3$ & $\rightarrow$ & $\begin{array}{ll} SO(2,1)_+ \times SO(2)_+^3 \\ SO(3)_+ \times SO(1,1)_-^2 \times SO(2)_+ \end{array}$ & $\begin{array}{c} \surd \\ - \end{array}$ \\\hline
\end{tabular}
\end{center}
\caption{Truncations of $\cN =4$ gaugings with de Sitter solutions having $SO(6,6)$ instabilities to $\cN = 2$ gaugings with $n_V = 5$, $n_H = 2$.} \label{tab:2442trunc}
\end{table}

\noindent
Subsequently, a careful analysis of the potential and its second derivatives can be carried out in order to determine whether the tachyonic scalars are truncated out. Using this procedure, we have been able to identify five stable de Sitter vacua in $\mathcal{N}=2$, that we will now list. 

For each specific truncation, we will indicate the extremum value of the potential, as well as the mass eigenvalues of the scalar fields (normalized by the potential) and their multiplicities (we will not list the mass eigenvalues of the $SL(2)$ scalars as these are positive in all cases). Whenever possible, we will explicitly indicate whether these masses are associated to vectormultiplet or hypermultiplet scalars. Note that this is only possible when the mass matrix splits up in two blocks, corresponding to vector- or hyperscalars respectively. As in \cite{Westra1,Westra2}, we use the notation $a_{ij} = g_i g_j \sin(\alpha_i - \alpha_j)$. The indices $i,j$ indicate the specific gauge factor of the $\cN = 4$ gauge group, in the order as they are written here. The $g_i$, $\alpha_i$ then denote the coupling constant, resp. $SL(2)$ angle of the $i$-th gauge factor.  

\mbox{}

\textbf{Stable de Sitter vacua with $n_V = 3$, $n_H = 4$:}

\begin{itemize}
\item $SO(2,1)_+^2 \times SO(2,1)_-^2 \rightarrow SO(2,1)_+ \times SO(2)_+$

One has to put the put the coupling constants of the two $SO(2,1)_-$ factors equal to zero for consistency (i.e. $g_3 = g_4 = 0$). The potential then reaches the value $V_0 = a_{12}$ at the extremum. The masses of the vector- and hypermultiplet scalars are given by:
\begin{eqnarray}  \label{totrunc2}
\frac{m^2_{\mathrm{vec}}}{V_0} = \left\{ 0 \ (2 \times)\,, \ 1 \ (2 \times)\right\} \,, \nonumber \\
\frac{m^2_{\mathrm{hyper}}}{V_0} = \left\{ 0 \ (8 \times)\,, \ 1 \ (8 \times)\right\} \,.
\end{eqnarray}
\end{itemize}

\mbox{}

\textbf{Stable de Sitter vacua with $n_V = 5$, $n_H = 2$:}

\begin{itemize}
\item $SO(3,1)_+ \times SO(3,1)_+ \rightarrow SO(2,1)^H_+ \times SO(3)^H_+$

In this case the value of the potential at the vacuum is given by $V_0 = 3 a_{12}$. The mass matrix is not block diagonal. Its eigenvalues are explicitly given by:
\begin{equation}
\frac{m^2}{V_0} = \left\{0 \ (8 \times)\,, \ \frac{2}{3} \ (2 \times)\,, \ \frac{4}{3} \ (6 \times)\right\} \,.
\end{equation}
As indicated in the example given above, this model was discussed in \cite{FTP} and we find perfect agreement with their results.
\item $SO(3,1)_+ \times SO(2,2)_\pm\rightarrow SO(2,1)_+ \times SO(3)^H_+$

In order for this truncation to be consistent, one has to put the coupling constant $g_3$ of $SO(2,1)_-$ equal to zero. The value of the potential at the extremum is then given by $V_0 = \sqrt{3} a_{12}$. The mass eigenvalues for vector and hypermultiplet scalars are given by
\begin{eqnarray} \label{totrunc1}
\frac{m^2_{\mathrm{vec}}}{V_0} & = & \left\{ 0 \ (2 \times)\,, \ 1 \ (6 \times) \right\}\,, \nonumber  \\
\frac{m^2_{\mathrm{hyper}}}{V_0} & = & \left\{ 0 \ (2 \times)\,, \ \frac{2}{3} \ (6 \times) \right\} \,.
\end{eqnarray} 
This truncation corresponds to the third model discussed in \cite{FTP}\footnote{More specifically, the $SO(2,1)_+ \times SO(3)^H_+$ gauging corresponds to the third model of \cite{FTP} with $r_0 = 0$, while the $SO(2,1)_+^H \times SO(3)^H_+$ gauging has $r_0=1$.}. Again, the value of the potential and the mass eigenvalues are in agreement with their results.
\item $SO(2,1)_+ \times SO(2,1)_-^2 \times SO(2,1)_+ \rightarrow SO(2,1)_+ \times SO(1,1)_+^2 \times SO(2)_+$ 

The value of the potential at the extremum is now given by $V_0 = a_{14}$. For the masses of the vector- and hypermultiplet scalars, one finds:
\begin{eqnarray}
\frac{m^2_{\mathrm{vec}}}{V_0} & = & \Big\{ 1 \ (2 \times)\,, \  0 \ (2 \times)\,, \ \left(1 + \frac{a_{24}^2}{a^2_{14}} + \frac{a_{12}^2}{a_{14}^2} \right)\ (2 \times)\,, \nonumber \\ & & \quad \left(1 + \frac{a_{34}^2}{a^2_{14}} + \frac{a_{13}^2}{a_{14}^2} \right)\ (2 \times)\Big\} \,, \nonumber \\
\frac{m^2_{\mathrm{hyper}}}{V_0} & = & \Bigg\{ 0 \ (2 \times)\,, \ \frac{1}{a_{14}^2} \left( a_{12}^2 + a_{24}^2 \right) \ (1 \times)\,, \ \ \frac{1}{a_{14}^2} \left( a_{13}^2 + a_{34}^2 \right) \ (1 \times)\,, \nonumber \\ & & \left(1 + \frac{a_{24}^2}{a^2_{14}} + \frac{a_{12}^2}{a_{14}^2} \right)\ (2 \times)\,, \ \left(1 + \frac{a_{34}^2}{a^2_{14}} + \frac{a_{13}^2}{a_{14}^2} \right)\ (2 \times)\Bigg\}\,. \label{fourth}
\end{eqnarray}

\item $SO(2,1)_+^3 \times SO(3)_+ \rightarrow SO(2,1)_+ \times SO(2)_+^3$

In this case, the value of the potential at the extremum is given by $V_0 = \sqrt{a_{23}^2 + a_{12}^2 + a_{13}^2}$. The mass matrix exhibits a split between vector- and hypermultiplet masses, leading to:
\begin{eqnarray}
\frac{m^2_{\mathrm{vec}}}{V_0} & = & \Big\{ 0 \ (2 \times)\,, \ \frac{1}{V_0^2} \left(a_{12}^2 + a_{13}^2 \right) \ (2 \times) \,, \ \frac{1}{V_0^2} \left(a_{12}^2 + a_{13}^2 + 2 a_{23} V_0 \right) \ (2 \times)\,, \nonumber \\ & &  \quad \frac{1}{V_0^2} \left(a_{12}^2 + a_{13}^2 - 2 a_{23} V_0 \right) \ (2 \times) \Big\} \,, \nonumber \\
\frac{m^2_{\mathrm{hyper}}}{V_0} & = & \Big\{ \frac{1}{V_0^2} \left(a_{13}^2 + a_{23}^2 \right) \ (4 \times) \,, \ \frac{1}{V_0^2} \left(a_{12}^2 + a_{23}^2 \right) \ (4 \times) \Big\} \,.
\end{eqnarray}
Stability of the vacuum is achieved for e.g.~$a_{23}=0$. This can be achieved while still having $V_0 >0$, by putting the relevant coupling constants or $SL(2)$ angles to appropriate values. Note in the latter case of $\sin(\alpha_2 - \alpha_3) = 0$ all but two eigenvalues are strictly positive.

\end{itemize}

\noindent
In the previous we have listed all eigenvalues of the scalar mass matrix. However, for every non-compact generator in the gauge group, there is always a flat direction in the scalar potential corresponding to a Goldstone boson \cite{FTP}. The associated scalar is being eaten up by the gauge vector in order to render it massive via the BEH effect. Due to the $SO(2,1)_+$ factors there are therefore always two non-physical vanishing eigenvalues in the vectorsector. Furthermore, in the fourth example there are two non-physical zero eigenvalues in the hypersector. Both the fourth and the fifth example are therefore fully stable, with all physical scalars having strictly positive mass eigenvalues. These are the first such examples in the presence of hypermultiplets\footnote{We thank Mario Trigiante for pointing this out to us.}.

A subsequent question could be whether these models allow for a truncation of the hypersector. In terms of $SO(6) \times SO(6)$, such a truncation would correspond to modding out with a $\mZ_2$ element whose first $SO(6)$ factor is identical to that of the element to go to $\cN = 2$, while the second $SO(6)$ factor is the identity $\mI_6$. It can be seen that such a subsequent truncation is only possible in the absence of any gaugings of non-compact isometries of the quaternionic-K\"{a}hler manifold, i.e.~in the absence of any $SO(1,1)$ or $SO(2,1)_{\rm H}$ factors. In this way the hypermultiplet truncation of \eqref{totrunc2} leads to the first model of \cite{FTP}. Similarly, truncating the hypersector of \eqref{totrunc1} leads to the second model of \cite{FTP}. Hence also Fayet-Iliopoulos parameters can be generated in this way in models without a hypersector. We have also checked that none of the unstable models of tables \ref{tab:2244trunc} and \ref{tab:2442trunc} become stable after a truncation of the hypersector.


\section{Discussion}

In this paper we have shown that all known extended supergravity models with de Sitter solutions are related via supersymmetry truncations. In particular, we have discussed relations between the $\cN = 8$ and $\cN =4$ models, and between the $\cN = 4$ and $\cN = 2$ models. A natural question concerns the relation between $\cN = 8$ and $\cN = 2$. As follows from the previous discussion, only one of the five models of section 4 can be obtained in this way. This is the fourth model with eigenvalues listed in \eqref{fourth}. When descending in supersymmetry, the gauge groups of this model are
 \begin{align}
  SO(4,4) \quad \rightarrow \quad SO(2,2)_\pm^2 \quad \rightarrow \quad 
  SO(2,1)_+ \times SO(1,1)_+^2 \times SO(2)_+ \,. \label{N=8to2}
 \end{align}
As the $\cN = 4$ model has the restriction $\tilde{g}_1 = \tilde{g}_2 = 0$, we should impose $a_{14} = a_{24} = a_{23} = a_{13}$ and $a_{12} = a_{34} = 0$ on the $\cN = 2$ side.

Let us discuss the crucial ingredients of the $\cN =2$ models with stable de Sitter vacua. It was already pointed out in \cite{FTP} that their three models have the following features in common. First of all, the gauge group is a direct product of a compact and a non-compact gauge factor, which have different $SL(2)$ angles. Moreover, the compact factor needs to act non-trivially on the hypersector. From a rather large survey of candidate models with stable de Sitter vacua (listed in tables \ref{tab:2244trunc} and \ref{tab:2442trunc}), we have found only two additional possibilities. These satisfy the requirements identified by \cite{FTP}. The new $n_{V} = 5$ models are generalisations of \cite{FTP} by having abelian in addition to non-abelian gauge factors. These are the first examples of fully stable de Sitter vacua in $\cN = 2$ theories with hypermultiplets.

From the survey one can also extract the effect of a non-trivial action of the non-compact factor on the hypersector. As mentioned before, for the compact factor this was absolutely crucial. In contrast, it turns out that the opposite conclusion can be drawn for the non-compact factor. Indeed, having a non-compact gauge factor that acts on the hypermultiplet sector has a ``destabilising" effect. This can e.g.~be seen from table 2, where the unique stable model becomes unstable when one replaces $SO(2,1)_+$ by $SO(2,1)_+^H$ in the gauge group. This holds for both the second and the last line, which differ in the representation in which the $SO(2,1)_+$ acts on the hypersector.

Points that merit further investigation include the following. First of all, we have not considered the most general possibility to obtain stable de Sitter vacua by truncations of $\cN=4$ theories. One could in principal also obtain gaugings of $\cN = 2$ theories with a different number of vector- and hypermultiplets than considered here. One could also start from different $\cN = 4$ theories (e.g. with more vectors) and explore whether they allow for de Sitter vacua that become stable upon truncation. 

A final question regards the possible higher-dimensional origin of the stable vacua. In this respect it is useful to note that most of the stable de Sitter vacua we found cannot be directly obtained by truncation of an $\cN = 8$ theory. It was shown in \cite{Hull:1988jw} that the non-compact $\cN = 8$ gaugings can be associated to solutions of 11-dimensional supergravity with non-compact internal spaces. Our analysis however suggests that one cannot directly use this mechanism to interpret most of the stable $\cN = 2$ vacua from a higher-dimensional viewpoint\footnote{It would be interesting to see the geometric counterpart of the supersymmetry truncation \eqref{N=8to2} on the non-compact internal space for the case that does follow from $\cN = 8$.}. As an intermediate step towards a better understanding of this, one might consider stable $\cN = 2$ vacua in five dimensions \cite{Gunaydin:2000xk,CS} and their relation to the four-dimensional ones \cite{Ogetbil}.


\section*{Acknowledgements}


We thank Mees de Roo, Giuseppe Dibitetto, Sudhakar Panda, Mario Trigiante and Thomas Van Riet for stimulating discussions. Furthermore, we are very grateful to Mario Trigiante for providing us with a copy of his Mathematica routine to check stability of the $\cN = 4$ gaugings. The work of DR is supported by a VIDI grant from the Netherlands Organisation for Scientific Research (NWO).

\providecommand{\href}[2]{#2}\begingroup\raggedright\endgroup

\end{document}